\documentclass{aastex62}

\usepackage[encapsulated]{CJK}
\usepackage{ucs}
\usepackage[utf8x]{inputenc}

\usepackage{threeparttable}
\usepackage{rotating} 
\received{March 26, 2018}
\accepted{June 27, 2018}
\submitjournal{ApJ}

\shorttitle{New Oe stars}
\shortauthors{Li et al.}

\begin{document}

\title{New Oe stars in LAMOST DR5}

\correspondingauthor{Guang-Wei Li}
\email{lgw@bao.ac.cn}

\author{Guang-Wei Li}

\affiliation{Key Laboratory for Optical Astronomy, National Astronomical Observatories, Chinese Academy of Sciences, Beijing 100101, China}

\author{Jian-Rong Shi}
\affil{Key Laboratory for Optical Astronomy, National Astronomical Observatories, Chinese Academy of Sciences, Beijing 100101, China}
\author{Brian Yanny}
\affil{Fermi National Accelerator Laboratory, Batavia, IL 60510, USA}
\author{Zhong-Rui Bai}
\affil{Key Laboratory for Optical Astronomy, National Astronomical Observatories, Chinese Academy of Sciences, Beijing 100101, China}
\author{Si-Cheng Yu}
\affil{Key Laboratory for Optical Astronomy, National Astronomical Observatories, Chinese Academy of Sciences, Beijing 100101, China}
\author{Yi-qiao Dong}
\affil{Key Laboratory for Optical Astronomy, National Astronomical Observatories, Chinese Academy of Sciences, Beijing 100101, China}
\author{Ya-Juan Lei}
\affil{Key Laboratory for Optical Astronomy, National Astronomical Observatories, Chinese Academy of Sciences, Beijing 100101, China}
\author{Hai-Long Yuan}
\affil{Key Laboratory for Optical Astronomy, National Astronomical Observatories, Chinese Academy of Sciences, Beijing 100101, China}
\author{Wei Zhang}
\affil{Key Laboratory for Optical Astronomy, National Astronomical Observatories, Chinese Academy of Sciences, Beijing 100101, China}
\author{Yong-Heng Zhao}
\affil{Key Laboratory for Optical Astronomy, National Astronomical Observatories, Chinese Academy of Sciences, Beijing 100101, China}

%
%
%
%
%


\begin{abstract}

Stars of spectral type Oe are very rare. To date, only 13 Oe stars have been identified 
within our Galaxy. In this paper, we present six new Oe stars and four new B0e stars found in 
LAMOST DR5.  Repeated  spectral observations of the same Oe stars show some emission 
line variability. The H$\beta$ emission of TYC 4801-17-1 shows rapid V/R variation. Phase lags 
in the V/R ratio of TYC 4801-17-1 spectra are also seen. We found the unusual O4.5 star RL 
128 is an Oe star with variable H$\alpha$ intensity and its Ca II triplet emission appears when 
H$\alpha$ emission reaches maximum intensity.  These newly identified early type Oe and B0e 
stars significantly increase the known sample.

\end{abstract}



\section{Introduction} 
\label{intro}
The classic Oe spectral stellar type was defined by \citet{con74} as O type spectra
showing emission in the hydrogen lines, but without N III $\lambda$4634-4640-4642 
or He II $\lambda$4686  emission features. Some Oe stars additionally show emission in 
their He I lines. Stars of type Oe are often characterized by rapid rotation and 
double-peaked Balmer emission lines.  Differences between the origins of P Cyg profile 
(from strong stellar winds, not a characteristic of Oe stars) and the double-peaked emission profile 
characteristic of the Oe spectral type was clearly noted in Fig. 11 of \citet{con74}.  
\par

Stars of type Oe are very rare \citep{neg04}. To date, there are only 13 Oe stars known in 
the Galaxy, reproduced here in Table \ref{oeknown}. The Galactic O Star Spectroscopic 
Survey (GOSSS) \citep{sot11,sot14,gol16} confirms an Oe/O ratio of $0.03\pm 
0.01$. For comparison, \citet{zor97} presents a mean Be/B ratio of at least 17\%, 
and a B1e/B1 ratio is even 34\%.  These statistics are consistent with Oe stars representing the 
hot end of the Be temperature distribution, with the mechanism, a rotating disk, being common 
to Be and Oe stars. The rarity of Oe stars can be attributed to the hot strong O star stellar 
winds blowing away the envelope around the star and in most cases preventing a formation 
a rotating disk, where characteristic (Oe and Be) double-peaked emission lines originate. 
However, \citet{vin09} suggested Oe stars earlier than O9.5 have a different origin from 
classic Be stars.
\par
Metallicity can affect the stellar wind, as stars with lower metallicity tend to have weaker 
winds. A stellar wind can take away angular momentum, so more metal-poor star with less wind 
rotates more rapidly. Thus, a metal-poor star can form a rotating disk more easily, which can be 
retained even by the much hotter star. \citet{gol16} presented about 30 
classical Oe stars in Small Magellanic Cloud (SMC). The Oe/O frequency in SMC is 
0.26 $\pm$ 0.04, much higher than that in the Galaxy.  There are also 4 known 
LMC Oe stars with types as hot as O6 and O7. One of them, star 77616 even shows 
double-peaked He II $\lambda$4686 emission \citep{gol16}.

\begin{table}\centering
\caption{Previously known Galactic Oe stars}
	\label{oeknown}

\begin{threeparttable}
		
	\begin{tabular}{lllll}
		\hline
		Name &R. A.(degree) & Decl.(degree) &Spectral Type & References\\
		\hline
                                                  & &                    & OBe            & \citet{con74} \\
HD~24\,534 (X Per)& 58.846151 & 31.045837 & B0 Ve        & \citet{neg04} \\
                                                    & &                    &O9.5: npe    & \citet{sot11} \\\hline
                                  & &     &  O6 V?[n]pe  & \citet{wal73}\\                  	
HD~39\,680 &88.686377 & 3.854734	                   &   O8.5 Ve     &\citet{neg04}\\
                              & &          & O6 V:[n]pe var  & \citet{sot11} \\\hline
                  
                                            & &                  &  OBe       & \citet{con74} \\                  
HD~45\,314& 96.815740 &	14.889224  & B0 IVe     & \citet{neg04} \\
                                           & &                   &  O9: npe     &  \citet{sot11} \\\hline

HD~46\,056 & 97.836912 & 4.834401 & O8 V(e)     & \citet{fro76}\\
                  & & & O8 Vn&   \citet{sot11} \\\hline
             & &     &  O8 V:pe     & \citet{mor55} \\
HD~60\,848 & 114.273880 &	16.904252  &  O9.5 IVe    & \citet{neg04} \\
              & &    &  O8 V:pe     & \citet{sot11} \\\hline
                  
               & & &  O9 V(e)   & \citet{con74} \\
  HD~149\,757 ($\zeta$ Oph)  &249.289741 & -10.567090 &  O9.5 IV   &\citet{neg04}\\
                    & &                        &  O9.2 IVnn &\citet{sot14}\\\hline
                                            
             & &          &  O7.5 IIIe & \citet{con74}\\
HD~155\,806 &	258.830199&-33.548421    &  O7.5 III((f))e & \citet{neg04}\\
                & &        &  O7.5 V((f))z(e) & \citet{sot14} \\\hline

HD~203\,064(68 Cyg)  &319.613257&	43.945967  & O8 V(e)  & \citet{con74}\\
            & &        &     O7.5 IIIn((f))  & \citet{sot11}\\\hline
HD~240\,234  &348.64358	&59.83678   & O9.7 IIIe &  \citet{neg04}\\\hline
HD~17\,520 B   &42.810071	&60.386118	 &  O9: Ve  & \citet{sot11} \\\hline
HD~93\,190   &161.081738 &	-59.283011    & O9.7: V:(n)e & \citet{sot14} \\\hline
HD~120\,678  &208.235085&	-62.720624   & O9.5 Ve & \citet{sot14} \\\hline
V441 Pup   &112.223241	&-26.108020	   & O5: Ve & \citet{mai16} \\\hline
\end{tabular}

\end{threeparttable}
\end{table}

Similar to classical Be stars, Oe stars often show significant emission line variability: 
\citet{rau15b} found that H$\alpha$, H$\beta$, and He I $\lambda$5876 of two original Oe 
stars HD~45\,314 and HD~60\,848 showed strong variations. For HD~60\,848, the variations 
of the equivalent widths of these lines are obviously asynchronous, while for HD~45\,314, the 
emission lines are highly asymmetric and display strong line profile variations, and in 2013, 
these lines even changed from double-peaked to single-peaked. 
HD~120\,678 also shows complex both light and spectral variations\citep{gam12}. 
The origins of these variations are still unknown. A summary 
of long-term V/R (the ratio of violet peak strength to the red peak strength) 
 variations can be found in \citet{riv13} and references therein.

\par
Earlier stellar types have stronger stellar winds, so most Oe stars are later O type stars. The 
spectral classification of an O star is based on the ratio of the equivalent width of He II 
$\lambda$4542 to He I $\lambda$4471 where later O types have a small ratio. 
This standard, however, is not reliable for the Oe class. \citet{fro76} noticed that 
He I  $\lambda$4471 of 
HD~39\,680 and HD~60\,848 may suffer from emission infilling, which result in higher EW(He 
II $\lambda$4542) / EW(He I $\lambda$4471), thus mimicking earlier spectral types. All the 
spectral types given in Table \ref{oeknown} are from the standard classification 
methodology -- some may be cooler than their spectral type suggests.
\par
Details of the formation of the rotating "decretion" (outwardly moving) disks in Oe (and Be)
stars remain complex and puzzling \citep{str31,zor16,zor17,lee91,osa86,riv01}. Undoubtedly, 
a larger Oe sample can help understand the nature of this type of emission-line star.
\par
Further identifying characteristics of Oe (and related Be) spectra to be kept in mind as 
we search the LAMOST DR5 sample for new members of the class:
\citet{por03} summarized: (1) Classical Be 
stars are non-supergiant (i.e. luminosity class V to III) B stars with line emission in Balmer 
lines (and possibly other lines). The Be phenomenon is complex: not all B stars with 
emission lines are classical Be stars (see their Table 1).  
The definition of a classical Be star as a
non-supergiant also extends to Oe stars.  (2) The emission lines in classical 
Be stars originate in a decretion disk due to the high rotational speed of the star. 
(3) Be stars have different types of variability associated with the star and the
disk.  More specifically, the disk can disappear from time to time. That is 
why Be stars as usually defined as objects that "have, or had at one time 
... Balmer lines in emission" \citep{col87}. 
One differentiates between a "Be spectrum" and a "Be star", as a Be star can have 
a Be spectrum permanently or only on occasions. 
\par
Table \ref{oeknown} has a heterogeneous origin, where two or more spectral types are 
assigned to one star which imply that possible disappearance of the disk, possible errors in the
classification or/and different part of a spectrum used. For example, $\zeta$ Oph only shows 
emission of H$\alpha$ line in \citet{con74} and does not appear as an Oe star in \citet{sot14}, 
where the spectrum does not cover the wavelength of H$\alpha$ line. However, even in 
\citet{neg04}, $\zeta$ Oph is still not an Oe star. In fact, its quiescence time is much longer than 
emission episode\citep{kam93}. HD~60\,848 is an O9.5 star in \citet{neg04} compared to O8 in 
other two references, which indicate that He I $\lambda$4471 emission from disk affect the 
spectral classification.

\par
Regarding O stars, there are several categories that include emission lines without being 
Oe stars. Indeed, Oe stars are a minority among O stars with emission lines. \citet{sot11,sot14} 
list the other types of O stars with emission lines: Ofc, Onfp, Of?p, early Of/WN, and O Iafpe. 
We also note that, as the resolution of the LAMOST spectrographs are typically R$\sim$1,800,
when classifying Oe stars we should distinguish between observed/strict/uncorrected spectral 
types and those corrected for the infilling of He I 4471, something that can be done only with
good-quality high-resolution spectroscopy (see e.g. the paragraph on V441 Pup
in \citep{mai16}. In spite of these complexities, we believe uncovering more samples of 
the Oe phenomenon is important toward further progress in understanding their origins.

\par

The paper is organized as follows: In Section 2, we introduce LAMOST and the selection 
method of the sample; new Oe and B0e stars found in LAMOST DR5 are presented in 
Section 3 and Section 4 respectively; Finally, conclusions are given in Section 5.

\section{Introduction to LAMOST and selection method of sample }

The Large Sky Area Multi-Object Fiber Spectroscopic Telescope 
(LAMOST, also called Guoshoujing Telescope) \citep{cui12, wang96, su04} is a special 
reflecting Schmidt telescope with $5^\circ$ field of view and effective aperture 3.6m - 4.9 m. 
It accommodates 4,000 fibers on its focal plane and can obtain nearly 4,000 spectra during one 
exposure. Its wavelength coverage is 3650\AA - 9000\AA with R $\sim$ 1,800. Each of 16 
spectrographs records images of 250 fibers on two 4K $\times$ 4K CCDs. As of 
the end of July 2017, more than 9 million spectra have been obtained (see \url{http://dr5.lamost.org/}). 
\par
We select candidate O type stars in LAMOST DR5 by using the O type spectral star criteria 
given by \citet{mai16}, then the Oe candidates stars are further selected by eye. We checked all 
2D and 1D LAMOST spectra of individual exposures of Oe stars and rejected bad pixels or 
pixels contaminated by cosmic rays or poorly subtracted sky lines.  All candidate Oe stars were 
previously cataloged as being of early type (usually late O or early B), though some of the 
previous observations date back decades. Some, but not all had emission noted, but none of 
our Oe (B0e) identified objects had previously been classified of the Oe (B0e) type.
\par
We note that the wavelengths in LAMOST spectra are all in vacuum, but to be coherent with 
the literature, lines are still named by their air wavelengths. The spectral classification 
methodology and luminosity criterion used in this paper are from \citet{sot11},  
\citet{sot14} and  \citet{mai16}.

\par

\section{New Oe stars in LAMOST DR5}
\label{secOes}

Details of spectra of the six stars that we newly identify from LAMOST spectra as meriting 
the Oe spectral class as given here. 

\begin{figure*}
       \includegraphics[width=180mm]{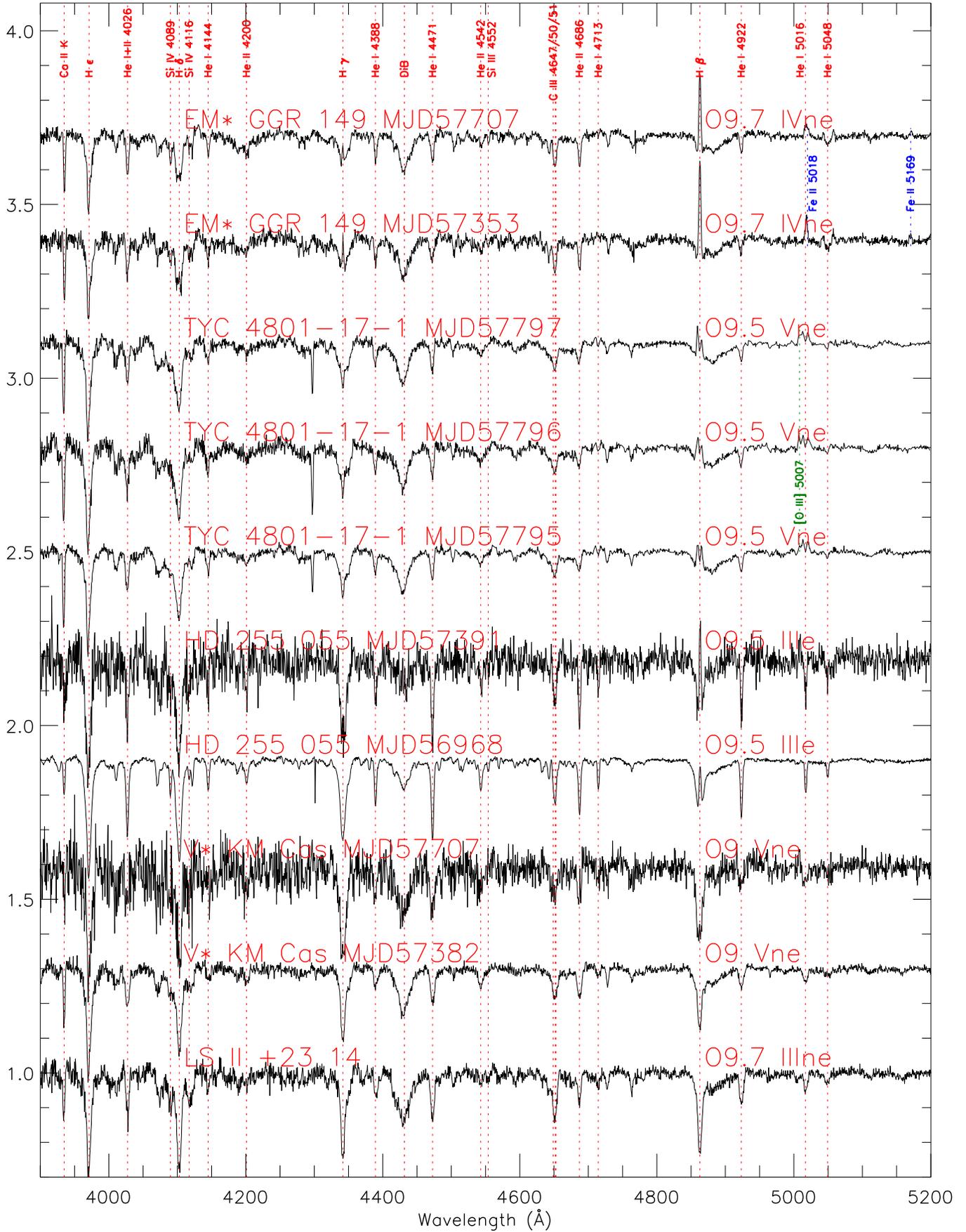}
    \caption{The blue spectra of Oe stars.}
    \label{oe}
\end{figure*}

\begin{figure}
	\includegraphics[width=18cm]{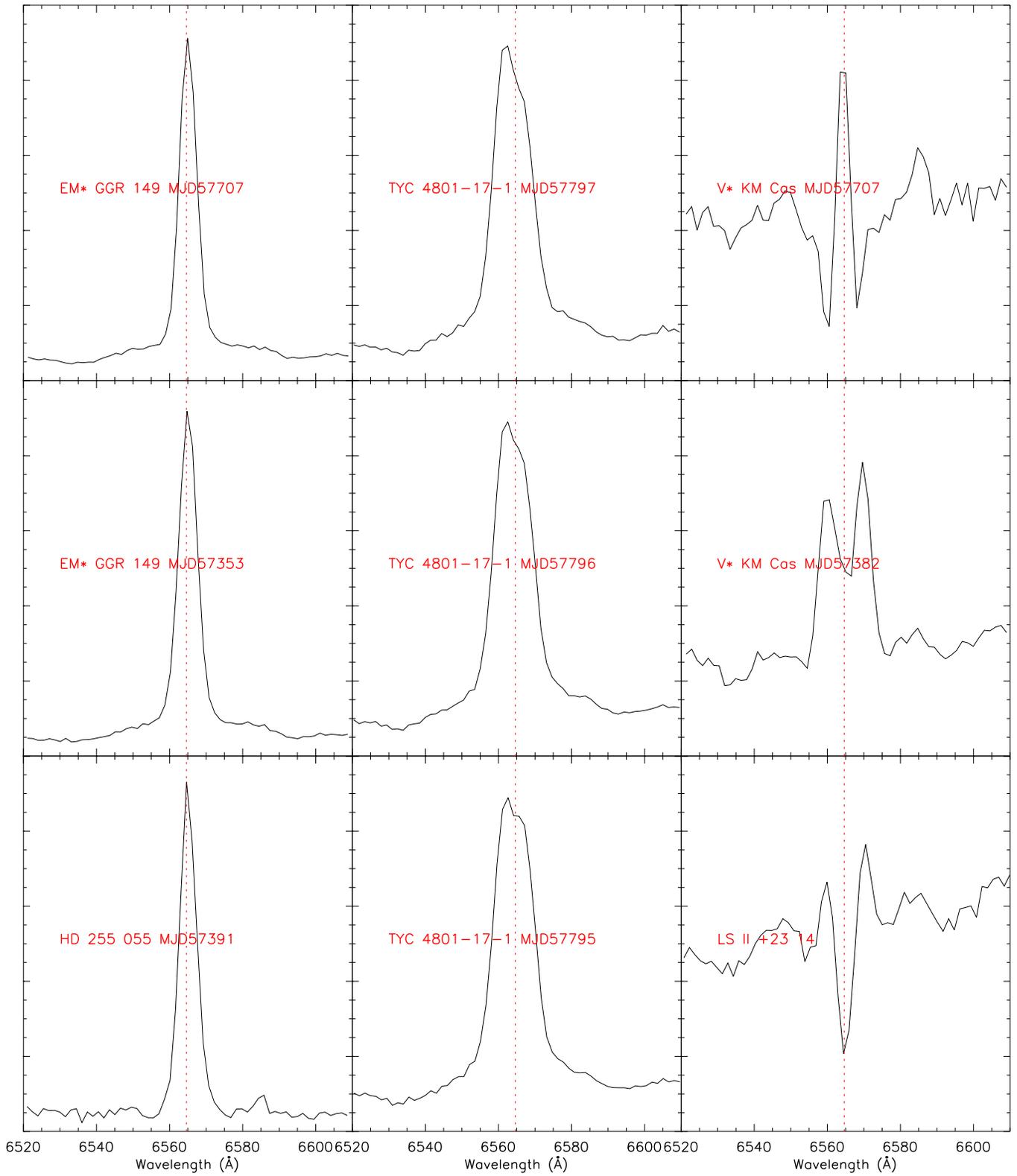}
    \caption{H$\alpha$ emission of selected Oe stars. The red line is the central wavelength 
    of H$\alpha$.}
    \label{oeha}
\end{figure}

\begin{figure*}
	\includegraphics[width=18cm]{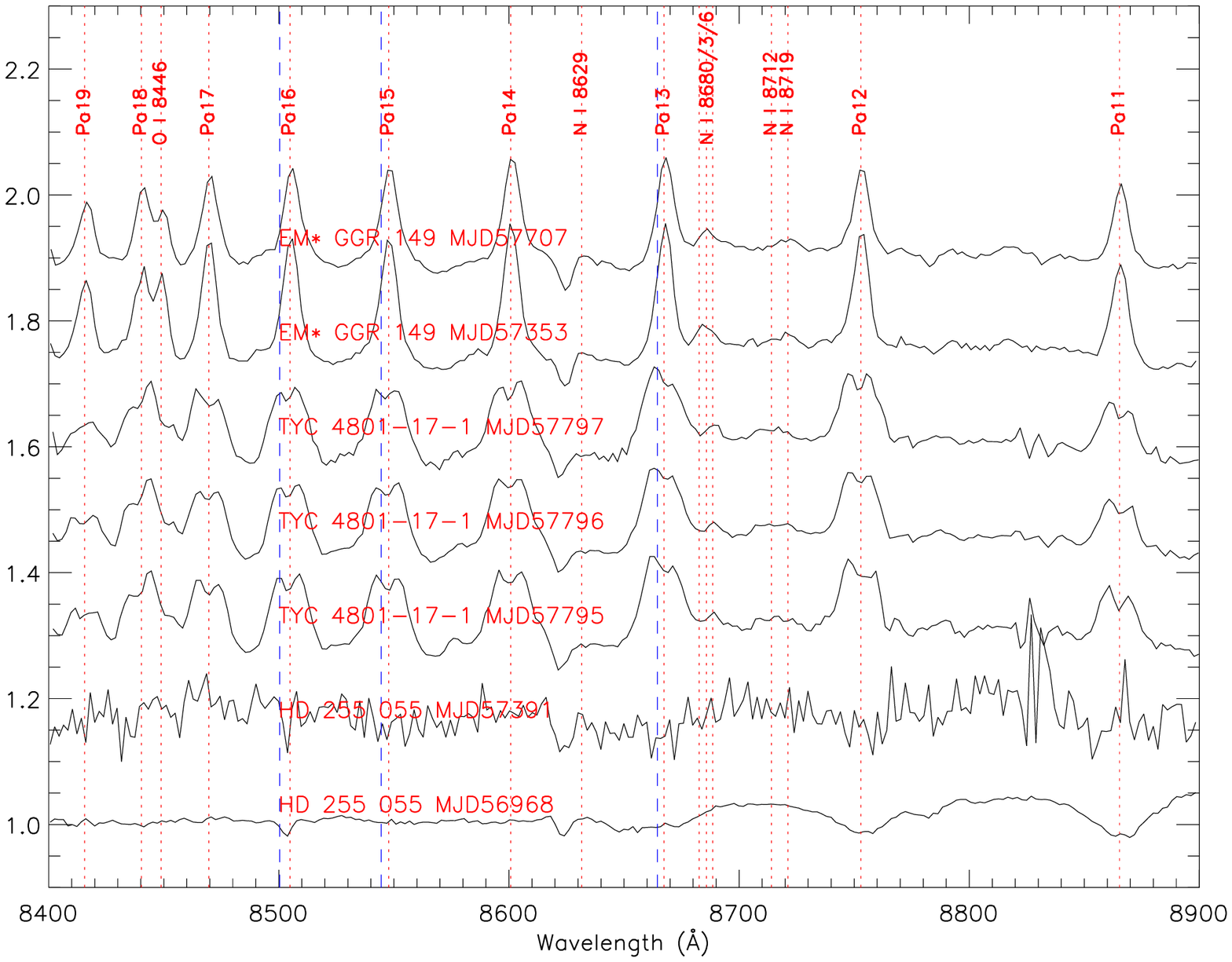}
    \caption{The z band spectra of selected Oe stars. The amplitudes of the emission lines 
    are suppressed for readability. Ca II triplets are indicated by the blue dotted lines.}
    \label{oepa}
\end{figure*}


\textbf{EM* GGR 149} In SIMBAD, it was assigned B0\citep{bro53,gon56} or 
OB\citep{koh99,wac70}. We found two spectra of this star in LAMOST DR5. In 
Fig. \ref{oe}, the ratios of He II $\lambda$4542 / Si III $\lambda$4552 indicate its 
spectral type should be O9.7. The ratios of He II $\lambda$4686 /  He I 
$\lambda$4713 indicate it is a dwarf, but the the Si IV $\lambda$4089 and $\lambda$4116 
lines imply its luminosity class IV. We notice that He I $\lambda$5048 has emission 
wings, so He I $\lambda$4713 may suffer from infilling. The round tips of He II 
$\lambda$4686 indicate it is a rapid rotator. Thus, its spectral type should be O9.7 IVn. 
\par
Emission from Balmer series lines through H$\epsilon$ and Fe II $\lambda$5018 and $\lambda$5169 
can be seen with intensity variations between two epochs, as shown in Fig.s \ref{oe} and 
\ref{oeha}. He I $\lambda$5048 shows double emission wings. Fig. \ref{oepa} illustrates its 
Paschen series, O I $\lambda$8446 and N I emission lines.

\par

\textbf{TYC 4801-17-1 = EM* RJHA 83} 
It H$\alpha$ emission was firstly reported by \citet{rob89}.We have three spectra in LAMOST
DR5. In these three spectra, He II $\lambda$4542 / He I $\lambda$4388 and He II 
$\lambda$4200 / He I $\lambda$4144 indicate its spectral type O9.5. We notice that He I 
$\lambda$4713 and $\lambda$5016 are double-peaked emission lines, but weak Si IV 
$\lambda$4089 and $\lambda$4116 lines imply it is a dwarf. The broad He lines indicate it 
is a rapid rotator. Thus, we assign it a spectral type O9.5 Vn. 

\par
In Fig. \ref{oe}, the most outstanding difference between these three spectra is the 
variation of the single-peaked [O III] $\lambda\lambda$5007 emission, while its neighbor, 
double-peaked He I $\lambda$5016, remains unchanged. 
\par
The double-peaked H$\alpha$ and H$\beta$ emission lines can be seen in Fig. \ref{oeha} 
and \ref{oe}, respectively, with V/Rs ascending from MJD57795 to MJD57797. Its 
double-peaked Paschen series emissions are shown in Fig. \ref{oepa}. The V/Rs of 
H$\alpha$ and H$\beta$ are obviously greater than 1, while V/Rs of Paschen lines are all 
around 1. The N I $\lambda$8680/3/6 emission are obvious, while N I $\lambda$8629, 
$\lambda$8712 and  $\lambda$8719 emission are weak.  The double-peaked O I 
$\lambda$8446 emission can be inferred from V/R $<$ 1 of Pa18 and the weak red 
component of O I $\lambda$8446 emission, while the V/R $>$ 1 of Pa13 indicates 
weak Ca II triplet emission.
\par
Long-term V/R variations of Be stars are well-known since \citet{mcl61} with a mean 
period of about 7 years. TYC 4801-17-1 shows clearly rapid V/R variations over three 
consecutive days. The dramatically ascending V/Rs of H$\alpha$ and H$\beta$ from 
MJD57795 to MJD57797 imply that TYC 4801-17-1 may have the rapidest V/R variation in 
known Oe class to date.
\par
The V/R phase lag is defined as different V/Rs between different emission lines in a 
spectrum. \citet{wis07} reported V/R phase lags between H$\alpha$ with V $<$ R and other 
emission lines with V $>$ R in $\zeta$ Tau. They supposed that there is a one-armed 
density wave with a significantly different average azimuthal morphology in the circumstellar 
disk, such that the different emission lines formed in different radii show different V/R 
phases, which is confirmed by \citet{ste09} and explained by a global disk oscillation 
model\citep{car09}. \citet{cho15} also presented the phase lags of Be stars in APOGGE 
spectra. TYC 4801-17-1 shows obvious V/R phase lags as have been mentioned above. 
Specifically, the V/R $\sim 1$ at Paschen series and He I $\lambda$5016, 
while the V/R $>1$ at H$\alpha$ and H$\beta$.
\par

\textbf{HD~255\,055}
In SIMBAD, it was assigned O\citep{wac70}, O9.0Vp(e?)\citep{cru74}, 
O9V:p\citep{hil56,mor55}, or O9V:pe:\citep{koh99}. There are five spectra in LAMOST DR5. 
Only two spectra with H$\beta$ at maximum and minimum phases respectively are shown in 
Fig. \ref{oe}. The star is a typical O9.5 star based on the ratios of He II 
$\lambda$4542/He I $\lambda$4388 and He II $\lambda$4200/He I $\lambda$4144 from 
the spectrum of MJD56968. The He II $\lambda$4686/He I $\lambda$4713 indicates 
luminosity class III. Thus, we assign it a spectral type O9.5 III. 
\par
Only the Balmer series lines show emission with variation -- shown in Fig. \ref{oe}. 
Its H$\alpha$ emission line is saturated in the spectrum of MJD56968, and not given in Fig. 
\ref{oeha}. The weak emission in the Paschen lines Pa11 and Pa12 can be seen in 
the spectrum of MJD56968 in Fig. \ref{oepa}, while other Paschen lines are infilled.
\par
It is also cataloged as a \textit{Kepler} K2 star (EPIC 202060093), showing photometric 
variation at low frequency without any cause given\citep{buy15}.

\par
\textbf{V* KM Cas}
It was classified as O9.5V((f)) by \citet{gon56} and \citet{mas95}.  There are two spectra in 
LAMOST DR5. In Fig. \ref{oe}, the ratios of He II $\lambda$4542/ He I $\lambda$4388  and 
He II $\lambda$4200/He I $\lambda$4144 are around 1, which indicate its spectral type O9, 
while its high ratio of He II $\lambda$4686/He I $\lambda$4713 indicates it should be a 
dwarf. The broad He lines indicate it is also rapid rotator. However, there is no sign of N III 
$\lambda$4634/40/41 emission. Thus, we assign it a spectral type O9 Vn.
In SIMBAD, it is an eruptive variable star, which implies that our spectra might miss the 
high phase of N III $\lambda$4634/40/41 emission. Its H$\alpha$ emission line shows variation 
between 2 epochs, as shown in Fig. \ref{oeha}.
\par

\textbf{LS II +23\,14} It was assigned to a spectral type B0Vn by \citet{ree03}. The spectrum 
in Fig.\ref{oe} indicates a spectral type O9.7 from the ratio of He II $\lambda$4542 / Si III 
$\lambda$4552, while the intensities of Si IV $\lambda$4089 and $\lambda$4116 lines 
indicate its luminosity III. The broad He I lines imply it is also a rapid rotator. Thus, it is an 
O9.7 IIIn star. Its H$\alpha$ in Fig. \ref{oeha} shows that it is an Oe shell star.
\par

\textbf{RL 128 = ALS~19\,265}        
The unusual properties of this star was noted by \citet{chr79}. It is very faint ($V = 15.11$), and very far from 
sun (65kpc) and was classified as O7. Recently, \citet{mai16} assigned it a spectral type 
O4.5V((c))z, based on their high quality spectrum. However, they also suspected that it may 
be a sdO. We found four spectra of RL 128 in LAMOST DR5. All our spectra confirm the spectral type O4.5 Vz, though the SNRs are all very low. RL 128 is also a 
\textit{Kepler} K2 star (EPIC 202060098) with light curve period 5.03 days \citep{buy15}, 
which was interpreted as rotation, but \citet{bal16} suggest a binary. The maximum 
velocity difference is about 20 km\,s$^{-1}$ between four LAMOST spectra, but it is still in 
question whether it is a binary or not because of low SNRs and resolutions of our spectra. 

\par
\begin{figure*}
	\includegraphics[width=180mm]{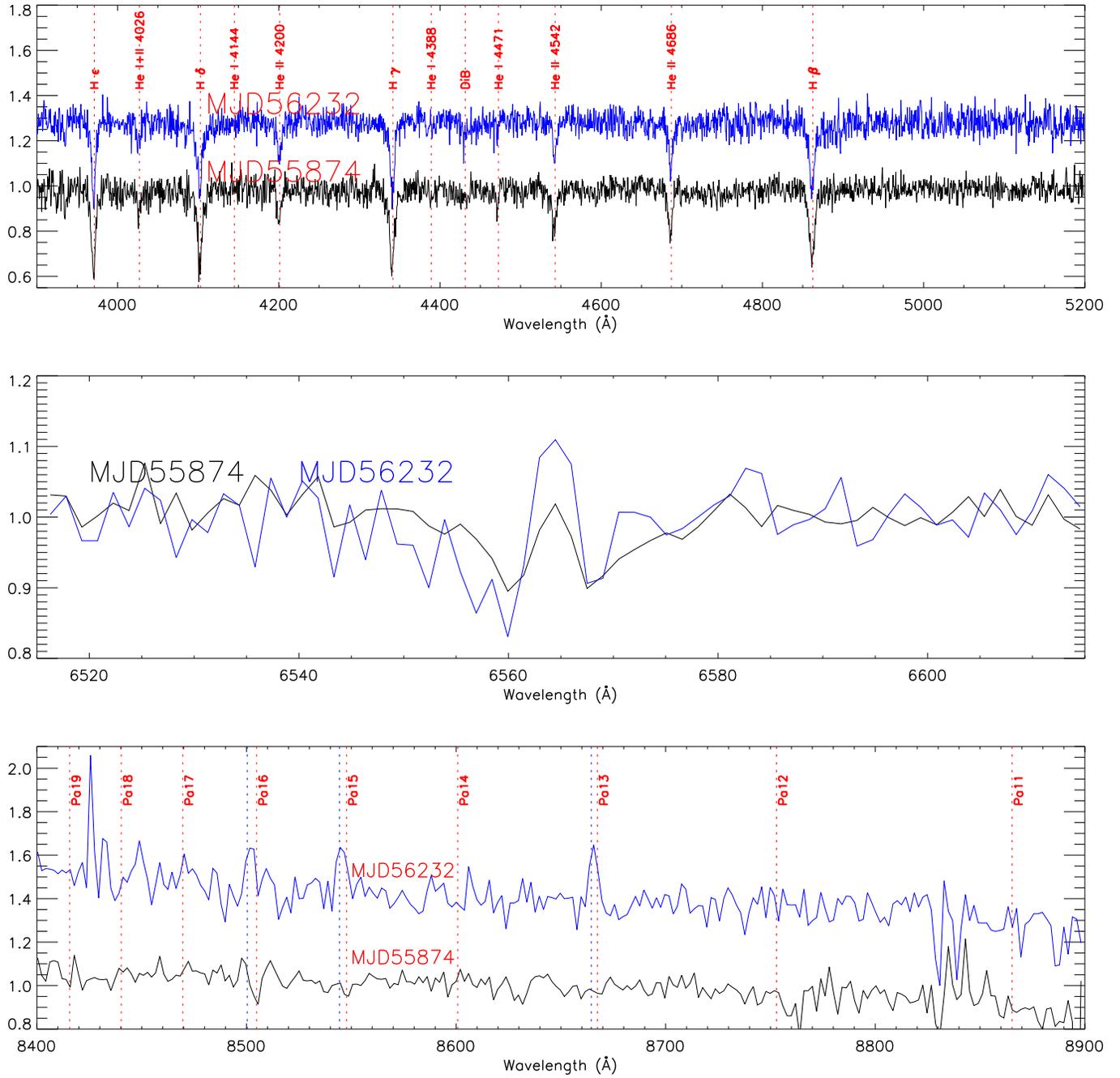}
    \caption{The two LAMOST spectra of star RL 128. The blue spectrum is for the maximum phase of the H$\alpha$ emission, while the black spectrum is for the minimum phase. The Ca II triplet lines are indicated by blue dashed lines.}
    \label{rl128}
\end{figure*}

For clarity, we only illustrate two spectra in Fig. \ref{rl128}, with H$\alpha$ reaching its 
maximum and minimum phases, respectively, among these four spectra. The Ca II triplet 
also show emission, though weak, when H$\alpha$ reaches its maximum intensity. 
The Oe character is based only on H$\alpha$, as no H$\beta$ emission is seen.

\section{New B0e stars in LAMOST DR5}
\begin{figure*}
	\includegraphics[width=180mm]{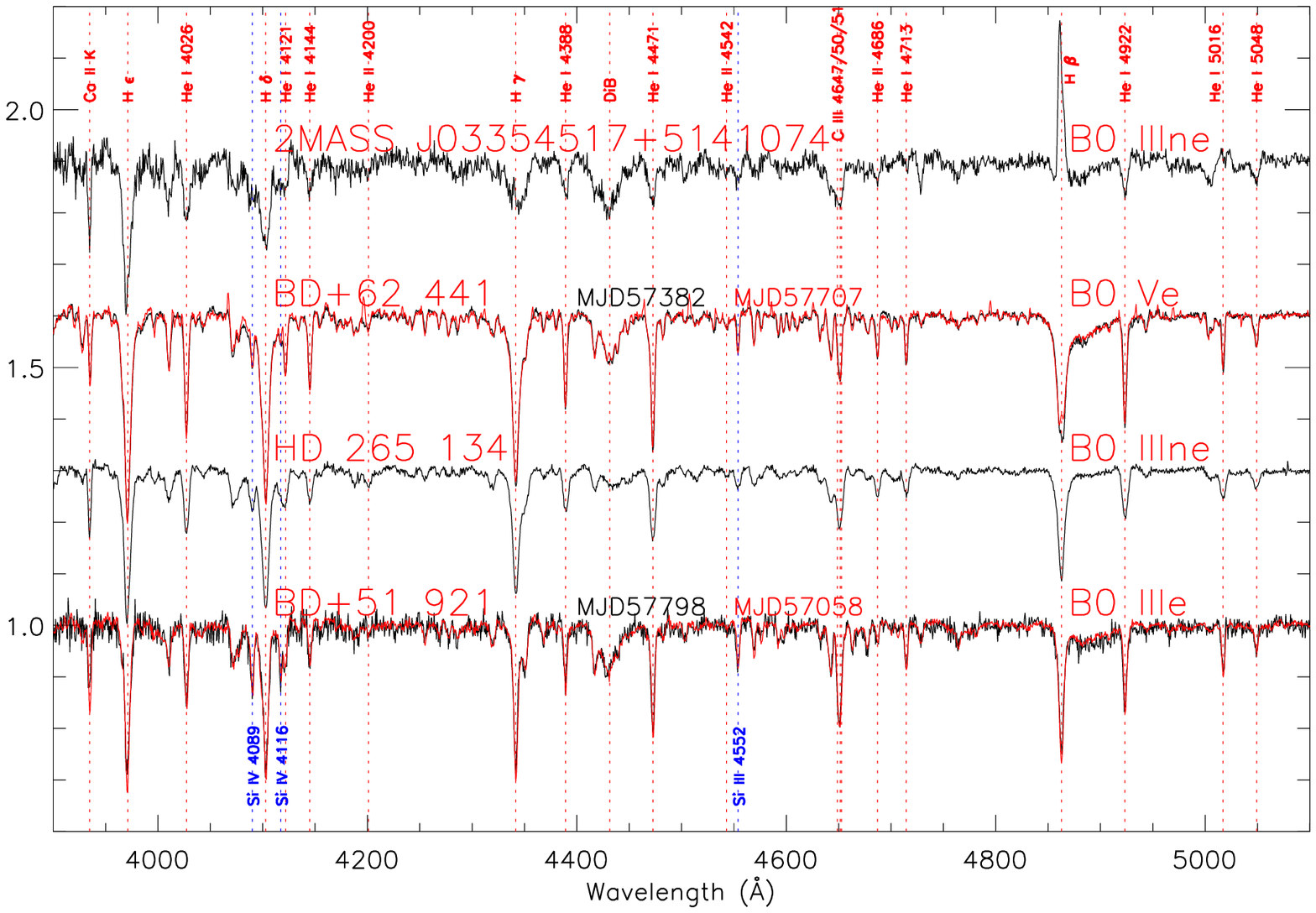}
    \caption{The blue spectra of B0e stars.}
    \label{b0e}
\end{figure*}

\begin{figure*}
	\includegraphics[width=12cm]{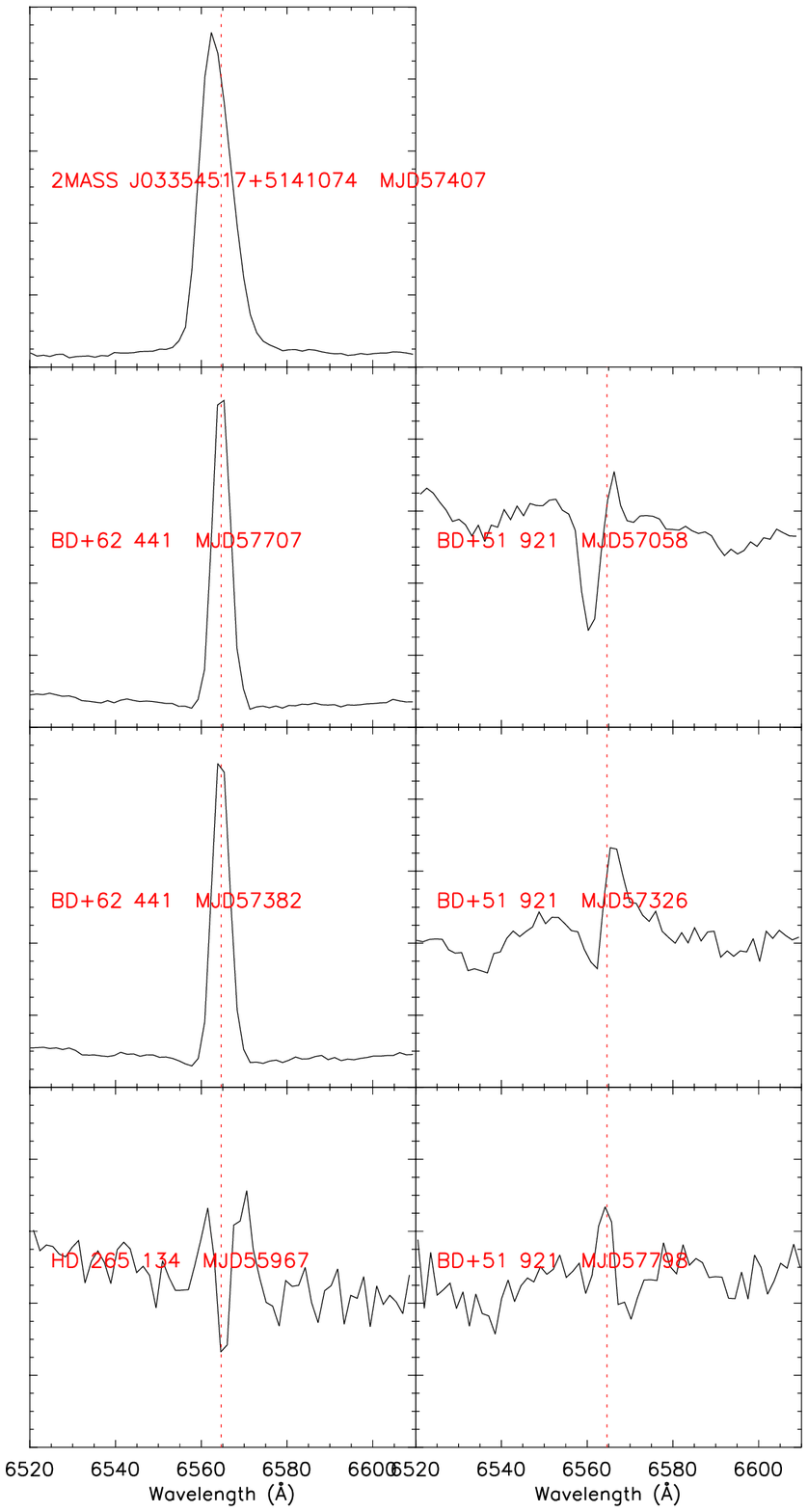}
    \caption{H$\alpha$ emissions of B0e stars.}
    \label{b0eha}
\end{figure*}

We find 4 new B0e stars in LAMOST DR5. The blue spectra of these B0e stars are 
shown in Fig. \ref{b0e}, while their H$\alpha$ lines are shown in Fig. \ref{b0eha}. 

\par

\textbf{2MSS\,J03354517+5141074}
We cannot find any information about this star in SIMBAD. In Fig. \ref{b0e}, its Balmer series until H$\delta$ are all shown emissions in their centers. 
Moreover, H$\beta$ emission is asymmetric. Besides, its He I $\lambda$5016, as well as 
$\lambda$5876, $\lambda$6678, $\lambda$7065, and $\lambda$7281 lines also show 
emissions. Its weak He II $\lambda$4542 indicates it is a B0 star, and its strong Si IV 
$\lambda$4089 and $\lambda$4116 lines indicate its luminosity III. Though suffering from 
infillings, its He I lines are all very broad, thus, we assign it a spectral type B0 IIIn. 
\par

\textbf{BD+62\,441}
There are two spectra of different epochs for this star in LAMOST DR5, which are 
overplotted in Fig. \ref{b0e} with the black spectrum from MJD57382 and the red spectrum 
from MJD57707. It was assigned B2III\citep{ryd78}, B0\citep{hec75} or B0I\citep{vor85}.  
The Weak Si IV $\lambda$4089 and $\lambda$4116 lines indicate its a B0 V star. Its 
H$\beta$ of MJD57707 is more emissive than that of MJD57382, though H$\beta$ infilling is weak in either epoch.
\par
\textbf{HD~265\,134} 
It was assigned B0 by \citet{nes95} or O9.5III by \citet{bis82}.  In Fig. \ref{b0e}, its weak He II 
$\lambda$4542 indicates it is a B0 star. The broad He I lines indicate it is a rapid rotator. 
Considering the shallow Si IV $\lambda$4089 and $\lambda$4116 are formed from the 
rotational broadening of intrinsically narrow deep lines, we assign it a spectral type B0 IIIn. 
Its H$\alpha$ shown in Fig. \ref{b0eha} indicates it is a Be shell star. No infilling is seen in H $\beta$ and the "e" suffix is based on H$\alpha$ only.
\par

\textbf{BD+51\,921}
There are three spectra from different epochs for this star in LAMOST DR5. Two of them 
are overplotted in Fig. \ref{b0e}, with the black spectrum from MJD57798 and the red 
spectrum from MJD57058. It was assigned to B0II by \citet{hil56}, but in Fig. \ref{b0e}, 
the similar intensity of Si IV $\lambda$4116 to that of He I $\lambda$4121 indicates  its 
luminosity class III, thus it is a B0 III star. The variation of H$\alpha$  profile between three 
epochs is obvious shown in Fig. \ref{b0eha}. No infilling is seen in H $\beta$ and the "e" suffix is based on H$\alpha$ only.

\par

\section{Conclusions} 
In this paper, we present 6 new Oe stars found in LAMOST DR5 data, increasing the 
numbers of known Oe stars by nearly 50\%.   We also list 4 new B0e stars found in LAMOST 
DR5. TYC 4801-17-1 shows rapid V/R variations at H$\beta$ and a V/R phase lag. 
Moreover, we find the unusual O4.5 star RL 128 is also an Oe star with variable H$\alpha$ 
intensity -- rare Ca II triplet emission appearsat the maximum phase of H$\alpha$ emission.
\par
All new Oe and B0e stars with their relevant spectral information from LAMOST DR5 are listed 
in Table \ref{oea} and \ref{b0ea}, respectively. The parallaxes and their errors are from Gaia 
DR2(see \url{https://gea.esac.esa.int/archive/}).

\acknowledgments

This research is supported by the National Natural Science Foundation of China (NFSC, 
Grant No. 11673036). We thank the expert anonymous referee, who provided generous 
detailed feedback that substantially improved the paper. 
\vspace{5mm}
\par
Guoshoujing Telescope (the Large Sky Area Multi-Object Fiber Spectroscopic Telescope 
LAMOST) is a National Major Scientific Project built by the Chinese Academy of Sciences. 
Funding for the project has been provided by the National Development and Reform 
Commission. LAMOST is operated and managed by the National Astronomical 
Observatories, Chinese Academy of Sciences.

\appendix


\begin{sidewaystable} 

\caption{New Oe stars in LAMOST DR5}
	\label{oea}

	\begin{tabular}{ccccccccccc} 
		\hline
 Star Name                  &  MJD    &  Plate ID            & Spectralgraph ID & Fiber ID &R. A.(degree) & Decl.(degree) & parallax (mas) & Spectral Type         \\\hline
  EM* GGR 149   &  56393    & Crab\_HIP115990\_2   &   09     & 080  &   353.810657	& 60.006443&  0.3166$\pm$   0.0249& O9.7 IVne  \\
                             &  57353    & Crab\_HIP115990\_2   &   09     & 080       &  \\
                             &  57707    & Crab\_HIP115990\_2   &   09     & 080       &   \\\hline
 TYC 4801-17-1   &  57795    & NGC230101          &   08     & 106       &103.183433	&	-0.187982& -0.8209 $\pm$  0.2103&O9.5 Vne \\
                 &  57796    & NGC230101          &   08     & 106       &    \\
                 &  57797    & NGC230101          &   08     & 106       &    \\\hline
   HD~255\,055  & 56968  & NGC2168\_3 & 06 & 210 &94.923531	&	23.288947& 0.5274 $\pm$0.0466&O9.5 IIIe \\
                &  57391    & KP061029N225952V01 &   13     & 040       &    \\
                & 57439  & KP062257N223048V01 & 15 & 232 &  \\
                 &  57442    &   kp2\_00\_6\_1       &   04     & 102       &    \\
                 &  57707    &   kp2\_00\_6\_1       &   04     & 102       &  \\
                 &  57707    &   kp2\_00\_6\_2       &   04     & 102       &  \\\hline
    V* KM Cas    &  57382    & NGC1027\_3          &   10     & 248       & 37.376939	& 61.495598&0.4063$\pm$ 0.0320&O9 Vne   \\
                         &  57707    & NGC1027\_3          &   10     & 248       &    \\\hline
  LS II +23 14   &  57884    & HD3385291          &   07     & 231       &295.051998 & 23.627886& 0.2725$\pm$ 0.0347&O9.7 IIIne \\\hline
 RL 128                &  55862   & B6212                 & 02        & 247       &96.249448 & 26.822039&0.1575$\pm$  0.0538& O4.5 Ve \\     
                             & 55874    &GAC\_097N28\_B1 & 02     &247 & \\
                             & 56232    &GAC097N26B1 & 03     &198 & \\
                             & 56639    &GAC097N26B1 & 03     &198 & \\\hline

\end{tabular}

\caption{New B0e stars in LAMOST DR5}
	\label{b0ea}

	\begin{tabular}{ccccccccc} 
		\hline
 Star Name                  &  MJD    &  Plate ID            & Spectralgraph ID & Fiber ID & R. A(degree)& Decl.(degree)& parallax (mas) &Spectral Type            \\\hline
 
2MASS J03354517+5141074  & 57407     & HD034854N505024V02 &   14  &  107 & 53.938267	&	51.685386&0.2851$\pm$ 0.0314& B0 Vne  \\\hline
 BD+62\,441  & 57382     & NGC1027\_3          &   15     &  023      &40.448588& 62.657267 & 0.3550$\pm$   0.0368&B0 Ve      \\         
                   & 57707     & NGC1027\_3          &   15     &  023      &          \\\hline
 HD~265\,134           & 55967     & GAC\_100N13\_V4      &   13     &  220      & 102.891369 & 13.617259 &0.1517 $\pm$  0.0833& B0 IIIne  \\\hline
 BD+51\,921           & 57058     & NGC1545\_3 &   12     &  249      & 66.362744	& 52.043780 &0.1416 $\pm$0.0358 &B0 IIIe      \\
                                & 57326     & HD042432N505042V01 &   15     &  137      &   \\
                                & 57798     & HD042432N505042V02 &   15     &  137      &  \\\hline

\end{tabular}

\end{sidewaystable}



\end{document}